\documentclass[12pt]{article}
\usepackage{amssymb}
\usepackage{amsmath}
\usepackage{epsfig}
\usepackage{float}
\newcommand{\be}{\begin{equation}}
\newcommand{\ee}{\end{equation}}
\newcommand{\bea}{\begin{eqnarray}}
\newcommand{\eea}{\end{eqnarray}}
\begin{document}

\begin{center}
{\bf STATUS OF NEUTRINO MASSES AND MIXING AND FUTURE PERSPECTIVES}
\footnote{Report at the conference IRGAC 2006, Barcelona July 11-15 2006}

\end{center}

\begin{center}
S. M. Bilenky
\end{center}
\begin{center}
{\em  Joint Institute
for Nuclear Research, Dubna, R-141980, Russia, and\\
Scuola Internazionale Superiore di Studi Avanzati,
I-34014 Trieste, Italy.}
\end{center}

\begin{abstract}
Status of the problem of neutrino masses, mixing and oscillations is discussed.
Future perspectives are briefly considered.
\end{abstract}

\section{Introduction}
Evidence for neutrino oscillations obtained in the Super-Kamiokande \cite{SK}, SNO \cite{SNO},
KamLAND \cite{Kamland}
and other neutrino experiments \cite{Cl}  is one of the most important recent
discovery in particle physics. There is no natural explanation of the
smallness of neutrino masses in the Standard Model.
A new, beyond the Standard Model mechanism of the generation of neutrino masses
is necessary in order to explain  experimental data.
In this talk we will discuss: I. Basics of neutrino mixing and neutrino oscillations. II. Evidence for neutrino oscillations. III. Open problems and future perspective.

\section{Basics of neutrino mixing and oscillations}

There are three basic ingredients in  the theory of neutrino
oscillations: neutrino interaction,  neutrino mass term, neutrino transition probabilities.

Neutrino interaction is well known. All existing experimental data
are perfectly described by Lagrangians of the charged current and neutral current
interactions of the Standard Model:
\begin{equation}\label{1}
\mathcal{L}_{I}^{\mathrm{CC}} = - \frac{g}{2\sqrt{2}} \,
j^{\mathrm{CC}}_{\alpha} \, W^{\alpha} + \mathrm{h.c.};~~
\mathcal{L}_{I}^{\mathrm{NC}} = - \frac{g}{2\cos\theta_{W}} \,
j^{\mathrm{NC}}_{\alpha} \, Z^{\alpha}.
\end{equation}
Here
\begin{equation}\label{2}
j^{\mathrm{CC}}_{\alpha} =2\, \sum_{l=e,\mu,\tau} \bar \nu_{lL}
\gamma_{\alpha}l_{L};~~~ j ^{\mathrm{NC}}_{\alpha}
=\sum_{l=e,\mu,\tau} \bar \nu_{lL}\gamma_{\alpha}\nu_{lL}
\end{equation}
are leptonic charged current and neutrino neutral current, $g$ is $SU(2)$ gauge constant and
$\theta_{W}$ is the weak angle.

Neutrino mass term determines neutrino masses, neutrino mixing and
nature of massive neutrinos. For neutrinos, particles with electric
charge equal to zero, there are two general possibilities for the
mass terms (see, for example, \cite{BGG})
\begin{center}
\bf{I. Majorana mass term}
\end{center}
Let us introduce column of the left-handed fields
\begin{eqnarray}
n_{L}=\left(
\begin{array}{c}
\nu_{e L}\\
\nu_{\mu L}\\
\nu_{\tau L}\\
\nu_{s_{1}L}\\
\vdots\\
\end{array}
\right) \label{3}
\end{eqnarray}
The fields $\nu_{s L}$ do not enter into the Lagrangians of the
standard CC and NC interaction (\ref{1}). By this reason they are
called sterile fields. The conjugated column
\begin{equation}\label{4}
(n_{L})= C~(\bar n_{L})^{T}~~ (C\,\gamma^{T}_{\alpha}\,C^{-1}= -
\gamma_{\alpha};\quad C^{T}=-C)
\end{equation}
is right-handed column. The most general Majorana mass term has the form
\begin{equation}\label{5}
\mathcal{L}^{\mathrm{M}}=-\frac{1}{2}\,\bar n_{ L}\,
M^{\mathrm{M}}\,(n_{ L})^{c} +\rm{h.c.}= -\frac{1}{2}\,\bar n_{ L}\,
M^{\mathrm{M}}\,C\,(\bar n_{ L})^{T}+\rm{h.c.}
\end{equation}
where  $M^{\mathrm{M}}$ is a symmetrical, complex matrix. We have
\begin{equation}\label{6}
M^{\mathrm{M}}=(U^{\dag})^{T}~m~U^{\dag},
\end{equation}
where $U$ is an unitary matrix and
$m_{ik}=m_{k}~\delta_{ik},~m_{k}>0 $. From (\ref{5}) and (\ref{6})
for the mass term we find
\begin{equation}\label{7}
\mathcal{L}^{\mathrm{M}}=-\frac{1}{2}\,\bar\nu ~m~ \nu
=-\frac{1}{2}\,\sum^{3+n_{s}}_{k=1}m_{k}~\bar\nu_{k}\nu_{k},
\end{equation}
where
\begin{equation}\label{8}
\nu=U^{\dag}~n_{L}+(U^{\dag}~(n_{L})^{c}= \left(
\begin{array}{c}
\nu_{1}\\
\nu_{2}\\
\nu_{3}\\
\vdots\\
\end{array}
\right)
\end{equation}
From (\ref{8}) we have
\begin{equation}\label{9}
    \nu^{c}_{i}= \nu_{i}~~( i=1,2,3,....3+n_{s}),
\end{equation}
where $n_{s}$ is the number of the sterile fields. Thus, {\em $\nu_{i}$ is the field of Majorana neutrinos with mass
$m_{i}$}. From (\ref{3}) and (\ref{8}) for the mixing we have
\begin{equation}\label{10}
\nu_{l L}=\,\sum^{3+n_{s}}_{k=1}U_{l k}\,\nu_{k L},\qquad\nu_{sL}=
\,\sum^{3+n_{s}}_{k=1}U_{s k}\,\nu_{k L},
\end{equation}
\begin{center}
\bf{II. Dirac mass term}
\end{center}
We will introduce now columns of {\em independent} left-handed and
right-handed fields
\begin{eqnarray}
n_{L}=\left(
\begin{array}{c}
\nu_{e L}\\
\nu_{\mu L}\\
\nu_{\tau L}\\
\nu_{s_{1}L}\\
\vdots\\
\end{array}
\right);~~~n_{R}=\left(
\begin{array}{c}
\nu_{e R}\\
\nu_{\mu R}\\
\nu_{\tau R}\\
\nu_{s_{1}R}\\
\vdots\\
\end{array}
\right) \label{11}
\end{eqnarray}
The most general Dirac neutrino mass term has the form
\begin{equation}\label{12}
\mathcal{L}^{\mathrm{D}}=-\bar n_{ L}\, M^{\mathrm{D}}\,n_{ R}
+\rm{h.c.}
\end{equation}
Complex matrix $M^{\mathrm{D}}$ can be presented in the form
\begin{equation}\label{13}
M^{\mathrm{D}}=V~m~U^{\dag},
\end{equation}
where $V$ and $U$ are unitary matrices and
$m_{ik}=m_{k}~\delta_{ik},~m_{k}>0 $. From (\ref{12}) and
(\ref{13}) for the Dirac mass term we find
\begin{equation}\label{14}
\mathcal{L}^{\mathrm{D}}=-\bar\nu ~m~ \nu
=-\sum^{3+n_{s}}_{k=1}m_{k}~\bar\nu_{k}\nu_{k},
\end{equation}
where
\begin{equation}\label{15}
\nu=U^{\dag}~n_{L}+V^{\dag}~n_{R}= \left(
\begin{array}{c}
\nu_{1}\\
\nu_{2}\\
\nu_{3}\\
\vdots\\
\end{array}
\right)
\end{equation}
From (\ref{14}) and (\ref{15}) it follows that $\nu_{k}(x)$ is the Dirac 
field of neutrinos ($L=1$) and antineutrinos ($L=-1$) ($L$ is the
conserved total lepton number). From (\ref{15}) for the 
neutrino mixing we will obtain Eq. (\ref{10}) in which $ \nu_{k}$ is
the field of the Dirac neutrino with mass $m_{k}$.

Type of the neutrino mass term at present is unknown. We will mention here only that 
in the case of the most popular see-saw mechanism of neutrino mass generation 
neutrino mass term is given by (\ref{5})  with $\nu_{s_{1}}=(\nu_{eR})^{c}$, 
$\nu_{s_{2}}=(\nu_{\mu R})^{c}$ and $\nu_{s_{3}}=(\nu_{\tau R})^{c}$. It is called 
Dirac and Majorana mass term and is a sum of left-handed and right-handed Majorana and Dirac mass terms for three left-handed and right-handed neutrino fields.
If left-handed Majorana matrix is equal to zero and right-handed Majorana matrix is much larger than 
Dirac mass matrix in this case massive neutrinos 
are Majorana particles with masses which are much smaller than masses of leptons or quarks.

We will consider now transitions of flavor neutrinos in vacuum. The
state of flavor neutrino $\nu_{l}$ with momentum $\vec{ p}$ produced
in a CC weak process together with $l^{+}$ is given by coherent
superposition of states of neutrinos with different masses
$|\nu_{i}\rangle$ (see \cite{BGG}):
\begin{equation}\label{16}
|\nu_{l}\rangle =\sum_{i}U_{l k}^*|\nu_{k}\rangle
\end{equation}
After time $t$ for neutrino state we have
\begin{equation}\label{17}
|\nu_{l}\rangle_{t} =\sum_{k}e^{-iE_{k}t}U_{l
k}^*|\nu_{k}\rangle=e^{-iE_{i}t}\sum^{3+n_{s}}_{k=1}e^{-i\frac{\Delta
m^{2}_{ik}}{2E}t}U_{l k}^*|\nu_{k}\rangle,
\end{equation}
where $\Delta m^{2}_{ik}=m^{2}_{k}-m^{2}_{i}$.
Neutrinos are detected via observation of CC and NC processes in
which flavor neutrinos $\nu_{l}$ take part. Existing neutrino
oscillation data  are described by the minimal scheme
of the three-neutrino mixing. In the case $n_{s}=0$ from (\ref{17})
we find
\begin{equation}\label{18}
|\nu_{l}\rangle_{t}=e^{-iE_{i}t}\sum_{l'}|\nu_{l'}\rangle~
\sum^{3}_{k=1}U_{l'i}~ e^{-i\frac{\Delta m^{2}_{ik}}{2E}t}\,U_{l
i}^*.
\end{equation}
From this expression for the probability of the transition
$\nu_{l}\to \nu_{l'}$ we find the following standard expression
\begin{equation}\label{19}
{\mathrm P}(\nu_{l} \to \nu_{l'}) =|\delta_{l'l}+\sum_{k=2,3}U_{l'
k} \,~ (e^{- i \Delta m^2_{1k} \frac {L} {2E}}-1) \,~U_{l k}^*\,
|^2,
\end{equation}
where $L\simeq t$ is the distance between neutrino production and
detection points.
Transition probabilities depend on six parameters (two neutrino
mass-squared differences $\Delta m^2_{12}$ and $\Delta m^2_{23}$,
three mixing angles $\theta_{12}$, $\theta_{23}$ and $\theta_{13}$
and one CP phase  $\delta$) and have rather complicated form.
However, two parameters are small:
\begin{equation}\label{20}
\frac{\Delta m^2_{12}}{\Delta m^2_{23}}\simeq
\frac{1}{30};~~~\sin^{2}\theta_{13}\leq 5\cdot 10^{-2}.
\end{equation}
{\em In the leading approximation in
which contribution of small parameters is neglected, neutrino
oscillations are described by simple two-neutrino expressions \cite{BGG}}. In
fact, let us consider atmospheric and long baseline accelerator
experiments with $\Delta m^2_{23} \frac {L} {2E}\gtrsim 1$
Neglecting contribution of small quantities $\Delta m^2_{12} \frac {L} {2E}$ and
$\sin^{2}\theta_{13}$ we can see that that dominant
oscillations in atmospheric region of $\frac {L} {E}$ are
$\nu_{\mu}\leftrightarrows \nu_{\tau}$. The probability of
$\nu_{\mu}$ to survive is given by the standard two-neutrino
expression
\begin{equation}\label{21}
{\mathrm P}(\nu_{\mu} \to \nu_{\mu})=1- {\mathrm P}(\nu_{\mu} \to
\nu_{\tau})=1-\frac {1} {2}\sin^{2}2\theta_{23}~(1-\cos\Delta
m^2_{23} \frac {L} {2E}),
\end{equation}
which depend only on two oscillation parameters
$\sin^{2}2\theta_{23}$ and $\Delta m^2_{23}$.

Let us consider now the reactor KamLAND experiment and 
solar neutrino experiments. For these experiments small $\Delta m^2_{12}$ is relevant
and the contribution to the transition probabilities of "large" $\Delta m^{2}_{23}$ is
averaged. Neglecting $\sin^{2}\theta_{13}$ for $\bar\nu_{e} \to \bar\nu_{e}$
survival probability in  vacuum (the  KamLAND experiment)  we find the following expression
\begin{equation}\label{22}
{\mathrm P}(\bar\nu_{e} \to \bar\nu_{e})=1-\frac {1}
{2}\sin^{2}2\theta_{12}(1-\cos\Delta m^2_{12} \frac {L} {2E}).
\end{equation}
For solar neutrinos MSW matter effect must be taken into account.
Neglecting the contribution of $\sin^{2}\theta_{13}$ we will find that  $\nu_{e} \to \nu_{e}$ 
transition probability is given by two-neutrino expression
\begin{equation}\label{23}
{\mathrm P}(\nu_{e} \to \nu_{e})= {\mathrm P}^{(1,2)}_{\rm{mat}}( \sin^{2}\theta_{12}, \Delta m^2_{12},
\rho_{e}),
\end{equation}
where $\rho_{e}$ is electron number density.
Thus,  in the leading approximation  {\em decoupling of oscillations in atmospheric-LBL  and
solar-KamLAND regions}
takes place. Existing experimental data are in agreement with such a picture of neutrino oscillations.

\section{Evidence for neutrino oscillations}

{\em First model independent evidence for neutrino oscillations was obtained in the 
Super-Kamiokande atmospheric neutrino experiment \cite{SK}.}  In this experiment atmospheric $\nu_{e}$ and 
$\nu_{\mu}$ are detected in the large 50 kt water Cherenkov detector. 
If there are no neutrino oscillations the number of electron and muon events must satisfy the following relation
\begin{equation}\label{24}
 N_{l}(\cos\theta)= N_{l}( -\cos \theta)\,~~ (l=e,\mu).
\end{equation}
Here  $\theta$ is zenith angle. Significant  violation  of this symmetry relation was observed in the case of  
high-energy muon events. For the ratio of the total number $U$
of up-going muons ($500 \lesssim L\lesssim 13000$ km) and total number
$D$ of  the down-going muons ($20 \lesssim L\lesssim 500$ km) it was found
\begin{equation}\label{25}
\left(\frac{U}{D}\right)_{\mu}= 0.551\pm 0.035 \pm 0.004
\end{equation}
In the Super-Kamiokande experiment 
$\frac{L}{E}$  dependence of the $\nu_{\mu}$ survival probability was measured.
From (\ref{21}) it follows that the survival probability has first minimum 
at $\Delta m^2_{23} \frac {L} {2E}= \pi. $ This minimum was clearly demonstrated by the data..
From analysis of the Super-Kamiokande data the 
following 90 \% CL
ranges of the oscillation parameters were found
\begin{equation}\label{26}
1.9\cdot 10^{-3}\leq \Delta m^{2}_{23} \leq 3.1\cdot
10^{-3}\rm{eV}^{2};~~~
\sin^{2}2 \theta_{23}> 0.9
\end{equation}
For the best-fit values of the parameters it was obtained:
\begin{equation}\label{27}
\Delta m^{2}_{23}=2.5\cdot 10^{-3}\rm{eV}^{2};~~\sin^{2}2 \theta_{23}=1.~~
(\chi^{2}/\rm{dof}= 839.7/755).
\end{equation}
The evidence for $\nu_{\mu}$ disappearance, obtained in 
the Super-Kamiokande atmospheric neutrino experiment, was  confirmed
by accelerator K2K \cite{K2K} and MINOS \cite{Minos} experiments. In the K2K experiment $\nu_{\mu}$'s produced at the 
KEK accelerator are detected by the Super-Kamiokande detector  at the distance of about  250 km. 
112 $\nu_{\mu}$ events were observed in the experiment.  In the case of no neutrino oscillations  
$158.1^{+9.2}_{-8.6}$ 
events were expected. From analysis of the K2K results for the best-fit values of the oscillation parameters
it was found the values
\begin{equation}\label{28}
\Delta m^{2}_{23}= 2.64\cdot 10^{-3}\rm{eV}^{2};~~\sin^{2}2\theta_{23}=1,
\end{equation}
which are compatible with the Super-Kamiokande 90 \% CL ranges (\ref{26}).

In the MINOS long-baseline accelerator experiment (Fermilab-Soudan, 730 km)
the number of expected  $\nu_{\mu}+\bar\nu_{\mu} $ events is $298\pm 15$. The number of  observed  events is equal to 204. From analysis of the data the following best-fit values of the oscillation parameters were obtained
\begin{equation}\label{29}
\Delta m^{2}_{23}=( 3.05^{0.60}_{-0.55} \pm 012) ~10^{-3}\rm{eV}^{2};~~~
\sin^{2}2\theta_{23}=0.88^{0.12}_{-0.15} \pm .
0.06
\end{equation}

In all solar neutrino experiments ( Homestake,
GALLEX-GNO, SAGE and Super-Kamiokande  \cite{Cl})
observed rates are 2-3 times smaller than the rates
predicted by the SSM\cite{SSM}.
{\em Model independent evidence for neutrino oscillations was
obtained in the solar SNO experiment \cite{SNO}}.
In this experiment solar neutrinos are detected via the
observation of three reactions
\begin{equation}\label{30}
\nu_e + d \to e^{-}+ p +p ~(\rm{CC});~
\nu_x+ d \to \nu_x + n +p~( \rm{NC});~
\nu_x  + e \to \nu_x + e~ (\rm{ES}).
\end{equation}
From the measurement of the CC rate the flux 
 of the solar $\nu_{e}$ on the earth can be inferred. From the measurement of the NC rate the
flux of all active neutrinos $\nu_{e}$, $\nu_{\mu}$ and $\nu_{\tau}$
can be determined. In the SNO experiment it was obtained
\begin{equation}\label{31}
\Phi_{\nu_{e}}^{\rm{SNO}} = (1.68 \pm 0.06 \pm 0.09) \cdot 10^{6}\,~
cm^{-2}s^{-1}; ~~\Phi_{\nu_{e,\mu,\tau}}^{\rm{SNO}} =(4.94 \pm 0.21 \pm 0.38
) \cdot 10^{6}\,~ cm^{-2}s^{-1}.
\end{equation}
The flux of $\nu_{e}$ on the earth is about three times smaller than the 
total flux of $\nu_{e}$, $\nu_{\mu}$ and $\nu_{\tau}$ :
\begin{equation}\label{32}
\frac{
\Phi_{\nu_{e}}^{\rm{SNO}}}{\Phi_{\nu_{e,\mu,\tau}}^{\rm{SNO}} }=
0.340 \pm 0.023 \pm 0.031
\end{equation}
Thus, it was proved by the SNO experiment that the solar $\nu_{e}$'s on the way from the sun to the earth
are transformed into $\nu_{\mu}$ and $\nu_{\tau}$.

The total flux of $\nu_{e}$, $\nu_{\mu}$ and $\nu_{\tau}$, measured by the SNO experiment, 
 is in agreement with the flux
predicted by SSM \cite{SSM}:
\begin{equation}\label{33}
\Phi_{\nu_{e}}^{\rm{SSM}}=(5.69 \pm 0.91  ) \cdot 10^{6}~
cm^{-2}s^{-1}
\end{equation}
{\em Model independent evidence for neutrino oscillations was found in the reactor 
KamLAND experiment.} In this experiments 
$\bar \nu_{e}$'s from  53 reactors in Japan are detected   via
the observation of the reaction $\bar \nu_{e}+p \to e^{+}+n$
by  a detector in the Kamiokande mine.
Average distance from reactors to the detector is about 170 km.  The experiment is 
sensitive to $ \Delta m^{2}_{12} $ and $\sin^{2} \theta_{12}$.
The expected  (without oscillations) number of the events in 
the KamLAND experiment is
$365.2 \pm
23.7$. The observed number of the events is  258. The ratio of the
observed and expected events is equal to $R= 0.658 \pm 0.044 \pm 0.047$.
Significant distortion of the spectrum of positrons produced in the
reaction $\bar\nu_{e}+p\to e^{+}+n $ was observed in the KamLAND
experiment.
From global analysis of  solar and KamLAND data for neutrino oscillation parameters it was found
the values 
\begin{equation}\label{36}
\Delta m^{2}_{12} = 8.0^{+0.6}_{-0.4}~10^{-5}~\rm{eV}^{2};~~~\tan^{2} \theta_{12}= 0.45^{+0.09}_{-0.07}.
\end{equation}
Summarizing, model independent evidence for neutrino oscillations
 driven by small neutrino masses and neutrino mixing were obtained in neutrino experiments.
All data (except LSND) are in agreement with the  three-neutrino mixing. Four neutrino oscillation parameters are 
determined from analysis of oscillation data with approximate accuracies:
\begin{equation}\label{37}
\Delta m^{2}_{12}~ (\sim 10\%); \tan^{2}; \theta_{12}~ (\sim 20\%) ;
\Delta m^{2}_{23} ~\rm{and}~ \sin^{2}2 \theta_{23} (\sim 30\%).
\end{equation}
For the parameter $\sin^{2} \theta_{13}$ only CHOOZ bound 
($\sin^{2} \theta_{13}\lesssim 5\cdot 10^{-2}$)  exists.  The CP phase $\delta$ is unknown.  
\section{Future perspectives}
For further progress the following basic questions must be answered:
\begin{center}
{\bf I. Are neutrinos with definite mass Majorana or Dirac particles?}
\end{center}
The most sensitive way to study  the  nature of
massive neutrinos  is the search for neutrinoless double $\beta$
decay 
\begin{equation}\label{38}
(\rm{A,Z}) \to (\rm{A,Z +2}) +e^- +e^-
\end{equation}
of $^{76} \rm{Ge} $, $^{130} \rm{Te} $, $^{136} \rm{Xe} $, $^{100}
\rm{Mo}$ and other even-even nuclei.
The matrix element of $0\nu\beta\beta$-decay is proportional to the
effective Majorana mass $m_{\beta\beta} = \sum_{i}U^{2}_{ei}\,m_{i}$.

From the best existing lower bounds on the half-lives of the $0\nu\beta\beta$-decay 
\bea
T^{0\nu}_{1/2}(^{76} \rm{Ge})\geq 1.9 \cdot 10^{25}\, \rm{years}~~ (\rm{Heidelberg-Moscow})\nonumber\\
T^{0\nu}_{1/2}(^{130} \rm{Te})\geq 5.5 \cdot 10^{23}\, \rm{years}~~ (\rm {Cuoricino})
\label{39}
\eea
for the effective Majorana mass the following bounds 
$|m_{\beta\beta}| \leq (0.2-1.2)\,~\rm{eV}$ can be inferred.
Many new  experiments on the search for
$0\nu\beta\beta$-decay (CUORE, GERDA,  EXO,  MAJORANA and others)
are now in preparation \cite{Elliot}. The goal of the future experiments is to reach the sensitivity 
$ |m_{\beta\beta}|\simeq \rm{a ~ few }\,~10^{-2}~
\rm{eV}$.
\begin{center}
{\bf II. What is the value of the lightest neutrino mass $ m_{0}$?}
\end{center}
From the measurement of the $\beta$-spectrum of $^{3}\rm{H}$ in Mainz and Troitsk experiments 
\cite{Mainz,Troitsk} it was obtained $m_{0}< 2.3$ eV. 
In the future KATRIN experiment \cite{Katrin} the sensitivity $m_{0}\simeq 0.2 $ eV
is planned to be reached.

From cosmological data for the sum of the neutrino masses it was found the range 
$\sum_{i}m_{i} <(02.-0.7)$ eV. The future data will be sensitive to 
$\sum_{i}m_{i}\simeq 5\cdot 10^{-2}$ eV (see \cite{Dodelson})
\begin{center}
{\bf II. What is the value of the parameter $\sin^{2}\theta_{13}$?}
\end{center}
If the value of this parameter is not too small it will be possible to study such fundamental 
effects of the three neutrino mixing as CP violation in the lepton sector and 
to reveal the character of neutrino mass spectrum.
The existing neutrino oscillation data are compatible with two types of neutrino mass spectra:
$m_{1}<  m_{2}    <  m_{3} ;~ \Delta m^{2}_{12}  \ll    \Delta
m^{2}_{23}  $ (normal spectrum) and 
$m_{3}<  m_{1}    <  m_{2} ;~ \Delta m^{2}_{12}  \ll
 | \Delta m^{2}_{13}|  $ (inverted spectrum).
New reactor experiments DOUBLE CHOOZ and Daya Bay \cite{2Chooz} are in
preparation at present. In these experiments, correspondingly,  factor 10 and 20  improvement in the sensitivity to $\sin
^{2}\theta_{13}$ is expected. In the accelerator T2K \cite{T2K} 
experiment  sensitivity to $\sin ^{2}\theta_{13}$ will be about 25
times better than in the CHOOZ experiment.
Future neutrino facilities (Super beam, $\beta$ -beam, Neutrino
factory \cite{Nfactory})  will allow to study phenomenon of neutrino oscillations
with very high  precision. 
There is no doubts that the accomplishment of this experimental program will be crucial for 
understanding of the  origin of neutrino masses and mixing.

I would like to acknowledge 
the Italian program
``Rientro dei cervelli'' for the support.

\end{document}